\documentclass[conference]{IEEEtran}
\IEEEoverridecommandlockouts

\usepackage{cite}
\usepackage{amsmath,amssymb,amsfonts}
\usepackage{algorithmic}
\usepackage{graphicx}
\usepackage{textcomp}
\usepackage{xcolor, soul}
\def\BibTeX{{\rm B\kern-.05em{\sc i\kern-.025em b}\kern-.08em
    T\kern-.1667em\lower.7ex\hbox{E}\kern-.125emX}}

\begin{document}

\title{Investigating 
catastrophic 
forgetting  \\ in sound event classification
\thanks{This work was funded by Jane and Aatos Erkko
Foundation, grant number 230048, ”Continual learning of sounds with deep
neural networks”. \\
We acknowledge CSC Finland for awarding this project access to the
LUMI supercomputer, owned by the EuroHPC Joint Undertaking, hosted
by CSC (Finland) and the LUMI consortium. }
}

\author{\IEEEauthorblockN{ Riccardo Casciotti, Annamaria Mesaros}
\IEEEauthorblockA{\textit{Signal Processing Research Center} \\
\textit{Tampere University}\\
Tampere, Finland \\
\{riccardo.casciotti, annamaria.mesaros\}@tuni.fi}

}

\maketitle

\begin{abstract}
This work investigates a number of approaches to prevent catastrophic forgetting in class incremental learning scenarios for sound event classification tasks. We analyze the problem using architectural and regularization approaches, using FSD50K and AudioSet datasets. We design incremental stages and solutions that selectively protect the kernels of the network from weight updates to prevent catastrophic forgetting, and a dynamic head solution that expands itself each time a new task is learned. The findings show that catastrophic forgetting mainly happens in deeper layers, in particular in the classifier head. For the studied in-domain sound classification problem, the solution that seems to alleviate catastrophic forgetting and is the most efficient is a full freezing of the feature extractor with a fine-tuning of the dynamic head classifier, showing little to no forgetting and great training stability, and a good balance between memory-stability and learning plasticity.
\end{abstract}


\section{Introduction}
Deep neural networks are a mathematical model thought to learn one objective function, not taking into account the potential need for sequentially learning additional tasks. For this reason, DNNs strongly suffer from the effects of catastrophic forgetting~\cite{ven_continual_2025,parisi_continual_2019}, which is the phenomenon characterized by the forgetting of past knowledge stored in a network whenever the network is used to learn new information. Incremental learning, also known as \textit{continual learning} and \textit{lifelong learning}, consists of having a model capable of learning multiple different tasks incrementally, without forgetting previous stored information~\cite{wang_comprehensive_2024,parisi_continual_2019}. Continual learning consists of a trade-off and an ideal balance between memory retention and learning plasticity~\cite{ven_continual_2025}. 

Continual learning is a fundamental capability for a network that aims to be robust to change and can evolve in time. Indeed, the incremental behavior can be reflected under different aspects; the model can learn a completely different task from the ones it knew before, or it can update the previous tasks because the more recent one has a domain shift, which requires a retraining to take that into account during inference. 

\begin{figure*}
    \centering
    \includegraphics[width=1\linewidth]{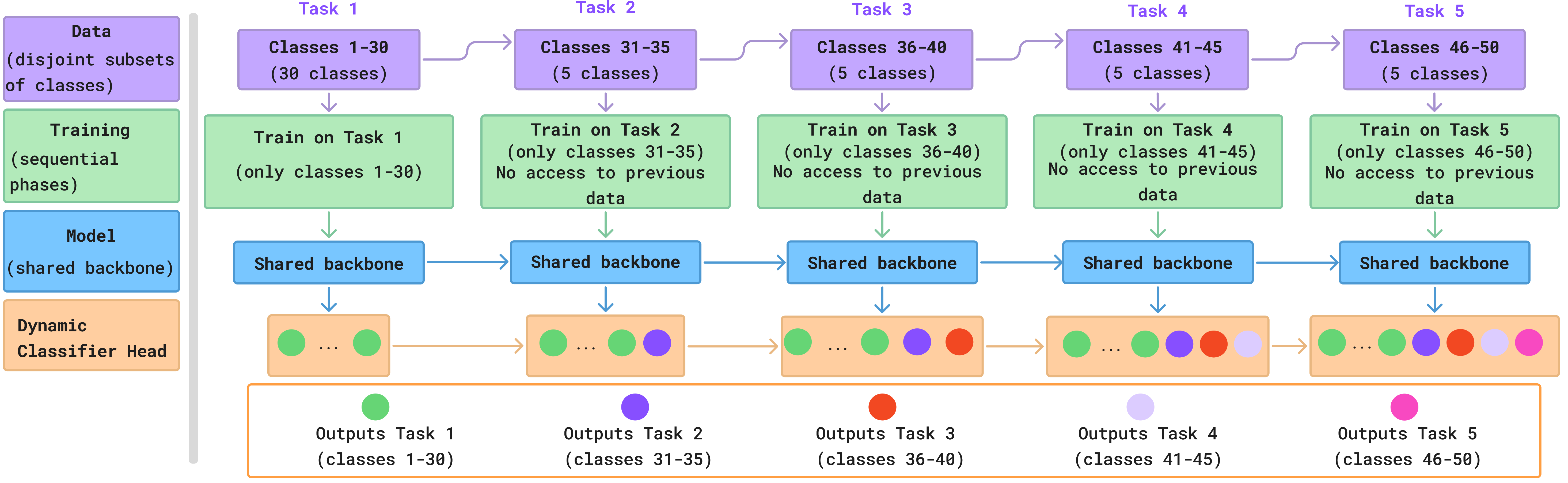}
    \caption{Incremental Learning setup with dynamic head expansion. Each task contains a disjoint set of classes to be learned, and the learning process has no access to the previous tasks data. The classification head expands at each task to accommodate the additional classes.}
    \label{fig:incremental_stages}
\end{figure*}

The specific continual learning scenario considered in this work is \textit{class-incremental learning}~\cite{van_de_ven_three_2022,hsu_re-evaluating_2019}. In class-incremental learning, the model receives data organized into a sequence of tasks, where each task introduces a new group of classes. The concept is illustrated in Fig.~\ref{fig:incremental_stages}. After training on a given task, the model is evaluated not only on the newly introduced classes, but also on all classes encountered so far. Therefore, the final objective is not simply to maximize performance on the current task, but to maintain a good balance between \textit{plasticity}, i.e. the ability to learn new classes, and \textit{stability}, i.e. the ability to retain previously acquired knowledge. This stability-plasticity trade-off is particularly relevant for sound event classification, where the input data are complex, noisy, and often multi-label. Unlike standard single-label classification, in audio tagging a single audio clip often contains multiple co-occurring sound events. This makes the learning problem more challenging, since the model must learn discriminative representations for overlapping acoustic patterns and preserve them across multiple incremental steps. In this work, we investigate continual learning for multi-class, multi-label sound event detection using two large-scale audio datasets, FSD50K and AudioSet~\cite{fonseca_fsd50k_2022,gemmeke_audio_2017}.

A wide range of continual learning strategies have been proposed in the literature~\cite{wang_comprehensive_2024,parisi_continual_2019}. One family of methods is based on regularization, where an additional loss term is introduced to constrain the learning process and prevent the model from drifting too far from previously learned solutions. Examples include methods based on knowledge distillation, where the outputs of a previous model are used as soft targets to preserve old knowledge~\cite{hinton_distilling_2015,li_learning_2017}, as well as parameter-importance approaches, where changes to important weights are penalized~\cite{kirkpatrick_overcoming_2017,zenke_continual_2017}. These methods act primarily through the optimization objective, modifying the loss function so that the final weight updates are indirectly regularized.

Another family of methods acts more directly on the optimization dynamics, for example by modifying gradient updates or parameter plasticity~\cite{he_gradient_2024,casciotti_incremental_2026}. Instead of only modifying the loss function, these approaches attempt to modulate the actual update received by individual parameters, filters, or kernels. This is particularly relevant when the goal is to protect specific parts of the network that are considered important for previously learned tasks. In this direction, one may estimate the importance of filters or kernels and then reduce or suppress their plasticity during future training. The intuition is that kernels that encode useful representations for previous classes should be protected, while less important kernels should remain available for adapting to new classes.

A related line of research concerns filter and kernel importance estimation~\cite{singh_efficient_2025}. In the pruning literature, many methods attempt to identify unimportant filters that can be removed from a network with minimal loss in performance. Although pruning and continual learning have different goals, they share a common requirement: estimating which parts of a network are important. If a pruning method can identify filters that are safe to remove, then its complementary information can be used to identify filters that should be preserved. In this work, we exploit this idea by using a passive filter pruning-inspired criterion to assign importance scores to kernels, then freezing different proportions of the most important kernels during incremental learning.

Finally, continual learning can also involve architectural modifications~\cite{rusu_progressive_2022,fernando_pathnet_2017}. In class-incremental learning, the output space grows as new classes are introduced~\cite{van_de_ven_three_2022,hsu_re-evaluating_2019}. In practice, this is often implemented by expanding the classifier layer to include output units for newly introduced classes. For example, if the first task contains 10 classes and the second task introduces 5 new classes, the classifier head is expanded from 10 to 15 output neurons. This modification is architectural in nature, although it is relatively lightweight compared to methods that add entire modules or task-specific branches. In our experiments, all methods use this classifier expansion mechanism, since the output space grows with each incremental task.

The main contribution of this paper is an experimental comparison of freezing, regularization, architectural expansion, and kernel-level plasticity modulation strategies under a class-incremental, multi-label sound event classification setting. We evaluate these approaches on FSD50K and AudioSet, using different task configurations and class orderings, and measure performance using both standard audio classification metrics and continual learning metrics.
Code is available on GitHub\footnote{https://github.com/RiccardoCasciotti/Class-Incremental-SEC.git} for experiment reproducibility.

\section{Methods}
In this work, we implement different continual learning approaches, which span from architectural solutions to regularization solutions. In particular, we focus on single kernel weights, and on the last fully connected layer, which for brevity is going to be referred to as the classifier head.\\

\subsection{Architectural methods}
Incremental learning requires the model to be able to learn tasks and maintain a good representation of them across numerous training phases. For this reason, we deemed it important to act architecturally on the model in a precise manner, acting on the classifier head. The classifier head can be considered as one of the most important components of a classifier model, because it is the layer that embeds the interpretation of the features extracted by the feature extractor. \\
The approach we took is to work with a dynamic output size classifier head. The dynamism is given by the ability of the classifier head to expand its outputs whenever the model is learning a new task. This setup gives it the possibility to associate a set of output heads specifically to each task, while keeping the outputs for the past tasks available. So the size of the classifier head increases with the increasing number of learned tasks. In particular, each class gets an output on the classifier head, as shown in Fig. \ref{fig:incremental_stages}.

The first approach we test is a knowledge distillation approach. The method is based on \cite{mulimani_class-incremental_2024}, where the effects of catastrophic forgetting are mitigated by considering a knowledge distillation loss and a cosine similarity loss. In this work, we only consider the knowledge distillation loss to protect past representations. This regularization solution acts on the loss function by adding a constraint that encourages the model trained on the new task to preserve the behavior of the model learned on previous tasks. 

In practice, before training on a new set of classes, a copy of the previous model is stored and used as a teacher. During the next training stage, the current model, acting as the student, is optimized not only to learn the labels of the new task, but also to reproduce the outputs produced by the teacher on the previously learned classes. These teacher outputs are usually treated as soft targets, because they contain more information than hard class labels: they encode the relative confidence of the old model across classes and therefore provide a compact representation of previously learned decision boundaries. The total objective is therefore composed of a standard classification loss for the new task and a distillation loss that penalizes deviations from the old model’s predictions. In this way, the method reduces catastrophic forgetting by discouraging the network from changing its output behavior too drastically on old classes while still allowing it to adapt to the new ones. Knowledge distillation is therefore a functional regularization method: instead of explicitly freezing parameters or selecting important kernels, it protects previous knowledge indirectly by preserving the input-output mapping learned by the model before the current incremental step.

The next approach uses a more aggressive solution, which regularizes the weight updates in the network during the training to try and protect the old knowledge the network has learned about previous tasks while learning a new set of classes effectively. The regularization method is based on \cite{singh_efficient_2025}, where the authors implement a passive filter pruning solution and use it to reduce the size of a deep neural network model, while keeping the performance almost unchanged. The core idea of \cite{singh_efficient_2025} is to estimate the importance of convolutional filters without using any data samples, unlike active pruning methods that require feature maps generated from a dataset. Instead of relying only on entry-wise filter norms, such as the l1- or l2-norm, the method evaluates filters according to their contribution to the output of the convolutional layer.

We use the method in \cite{singh_efficient_2025} to select kernels relevant for each task, thus allowing us to know which kernels should be protected from future updates. The solution is applied during the training phase of the network, where we calculate the vectors containing the ranks associated with each kernel in each layer, and  store them externally to the model after the training on a set of classes is completed. 
Whenever the model is resumed to learn a new set of classes (i.e. a new task), we also load the vectors containing the scores associated with each kernel, and  sort them in descending order. The kernels at the top of the list are considered the most relevant and important from previous tasks. During the training on a new set of classes, we protect those important kernels from incoming weight updates that would overwrite the information they contain, and we steer the learning towards less important kernels. We experiment with protecting a varying percentage of important kernels per layer: 12.5\%, 25\%, 50\%, 75\% and 100\%.
More details on this approach will be given in the experimental results in Section \ref{exp_setup}. 

A third approach we evaluate is based on \cite{casciotti_incremental_2026}; the work in \cite{casciotti_incremental_2026} utilizes a Hebbian deep neural network to implement a kernel-level plasticity modulation approach, which aims to protect old information while promoting learning plasticity for learning new tasks. In this paper, the kernel modulation solution used is implemented to work with backpropagation. The solution works as follows: during training on each task, the method  monitors how much each kernel changes by periodically storing its weights and computing the average magnitude of the weight updates. In parallel, it measures the cumulative activation of each kernel and ranks kernels according to how strongly they respond to the current task. The top-$k$ most active kernels are then treated as important for the task and stored. When the model later learns a new task, incoming updates are modulated according to these stored statistics: if an important kernel receives an update larger than its previously observed average update, its plasticity is reduced in order to protect the representation it has learned; at the same time, less-critical kernels, can receive increased plasticity so that adaptation to the new task is redirected toward parts of the network that are less important for previous tasks. In this way, the method attempts to preserve task-relevant kernels while still leaving enough capacity for learning new information. In the following we refer to this method as \textit{learning-modulation}.

Another method we implemented uses a more aggressive approach: a full freezing of all the convolutional layers, including the batch normalization layers, after the first task has been learned; this allows updating only the classification head for the incremental tasks, and is referred to in the following as \textit{fine-tuning}.

\section{Data and evaluation}
\label{data_section}
The experiments use two different datasets: AudioSet \cite{gemmeke_audio_2017} and FSD50K \cite{fonseca_fsd50k_2022}.  
AudioSet is a large-scale dataset of human-labeled audio events extracted from YouTube videos. It contains approximately 2.1 million 10-second audio segments annotated with one or more labels from a hierarchical ontology of 527 audio event classes, covering a wide range of real-world sounds. 
FSD50K is a large-scale, open dataset for sound event recognition composed of 51,197 audio clips collected from Freesound, annotated using 200 classes derived from the AudioSet ontology.

\subsection{Data preprocessing and experimental setup}

Both datasets are highly unbalanced: some classes have many thousands of samples while others have just a few hundred or even fewer. We selected the largest 50 classes of each set by number of samples to ensure we have enough training data for the  study. The 50 classes chosen will then be divided into disjoint sets for the incremental tasks.

The data is preprocessed to ensure all samples have the exact same length of 10 seconds: the shorter samples are padded with zeros, while the longer samples are cut into shorter ones of the desired length; log mel spectrogram is calculated as an input feature to the model. 
For both datasets, the parameters used to calculate the log-mel spectrogram are: 32000 as sampling rate (Hz), an FFT size of 1024, a hop length of 320 and 64 mel bands.

\subsection{Evaluation metrics}

The main metric used to measure the system performance is the Mean Average Precision (mAP), suitable for measuring the performance of the model in a multi-label setting. This metric is used during the training phase and the validation phase, in conjunction with a binary cross-entropy loss \cite{goodfellow_ian_and_bengio_yoshua_and_courville_aaron_deep_2016}. 

To evaluate the continual learning process, we use specific continual learning metrics that reflect how the model behaves across different tasks and while learning new classes. The CL metrics used are: Backward Transfer, Intransigence Measure, and Forgetting Measure. 

\textbf{Forgetting Measure}
\cite{lopez-paz_gradient_2017} (FM) directly quantifies catastrophic forgetting on previous tasks when learning a new one, defined as:
\vspace{-2pt}
\begin{equation}
\text{FM}_{k} = \frac{1}{k-1} \sum_{j=1}^{k-1} f_{j,k}\label{eq:7} \\
\end{equation}
%\vspace{-2pt}
where 
$f_{j,k} = \max_{i \in \{1,\dots,k-1\}} mAP_{i,j} - mAP_{k,j}, \forall j < k $ 
and $mAP_{i,j}$ is the mean average precision on task $i$ after learning task $j$.

\textbf{Backward Transfer} 
\cite{ferrari_riemannian_2018} (BWT) measures the interference of new tasks on previous ones and is defined as:
\vspace{-2pt}
\begin{equation}
\text{BWT}_{k} = \frac{1}{k-1} \sum_{j=1}^{k-1} mAP_{k,j} - mAP_{j,j} \label{eq:8} \\
\end{equation}
\vspace{-2pt}

\textbf{Intransigence Measure} 
\cite{ferrari_riemannian_2018} (IM) measures the difference in performance between the model in the incremental learning scenario and the joint model:
\vspace{-2pt}
\begin{equation}
\text{IM}_{k} = mAP^{*}_{k} - mAP_{k,k} \label{eq:9}
\end{equation}
\vspace{-2pt}where $mAP^{*}_{k}$ is the mAP of a reference model trained jointly on task $k$, and $mAP_{k,k}$ is the mAP on task $k$ right after training.

\section{Experimental Setup}
\label{exp_setup}
The experiments simulate an incremental learning scenario where a model has to learn five tasks in different training phases. Then the model is evaluated on the performance it has reached on each task and on the information it remembers from previous training steps. 

The 5 tasks are organized by generating five disjoint subsets of classes from the same dataset, where the first task has 30 classes, and the 4 following tasks have 5 classes each. The model learns the first set of classes; then, without seeing the old classes again, it is trained on the new subset of classes through a separate training procedure, and so on until all five sets of classes are learned. Each set of classes corresponds to a new multi-label classification task that the model has to learn how to solve. 

The experiments are run using two different configurations, for both FSD50k and AudioSet,  
one learning the classes in order of their size, and the other one in random order. 
For brevity, we are going to refer to the first configuration as "ordered" and the second configuration as "not ordered". 
The main difference between the two experimental setups is that in the former the first task will contain the classes with the highest amount of samples compared to the incremental tasks. Differently, in the latter setup a class with a very high number of samples could also appear in a later task. We investigate both scenarios to observe any potential differences between methods that are an effect of class learning order. The random selection of classes into the five incremental tasks was performed 5 times. The results presented in Section \ref{sec:results} are an average over all these runs.

\subsection{Model Architecture}
The architecture is a PANNs-style convolutional neural network \cite{kong_panns_2020}. We chose to adopt this model because of its proven performance 
in many sound classification tasks.
Our architecture consists of six convolutional blocks each containing two convolutional layers followed by a batch normalization layer, and followed by a fully connected layer which serves as the dynamic classifier head. Since the task addressed in this work is a multi-label classification problem, the activation function applied to the final layer is sigmoid. 

\subsection{Training and validation}
We trained our models using the Stochastic Gradient Descent optimizer \cite{Bottou2012} using a callback to reduce the learning rate with a cosine annealing schedule and early stopping with a patience of 20 epochs, and a starting learning rate of 0.001.
The model was trained using a binary cross-entropy loss computed on the classes of the current task.

The tasks are composed of disjoint subsets of classes of the same dataset, chosen as described earlier. 
The training is organized as follows: a freshly initialized model is trained on the first task comprising 30 classes, then the same model undergoes a dynamic change by increasing the number of output heads in the last layer by 5. After this, the model starts the training phase for the second task, without access to the data of the previous task(s), and so on until the last task is reached. 
A validation phase is run right after the training of each task, using a dedicated validation set extracted from each dataset. The validation dataset contains only current task data.

\section{Results and discussion}
\label{sec:results}
\begin{figure*}%[!t]
\centerline{\includegraphics[width=\textwidth]{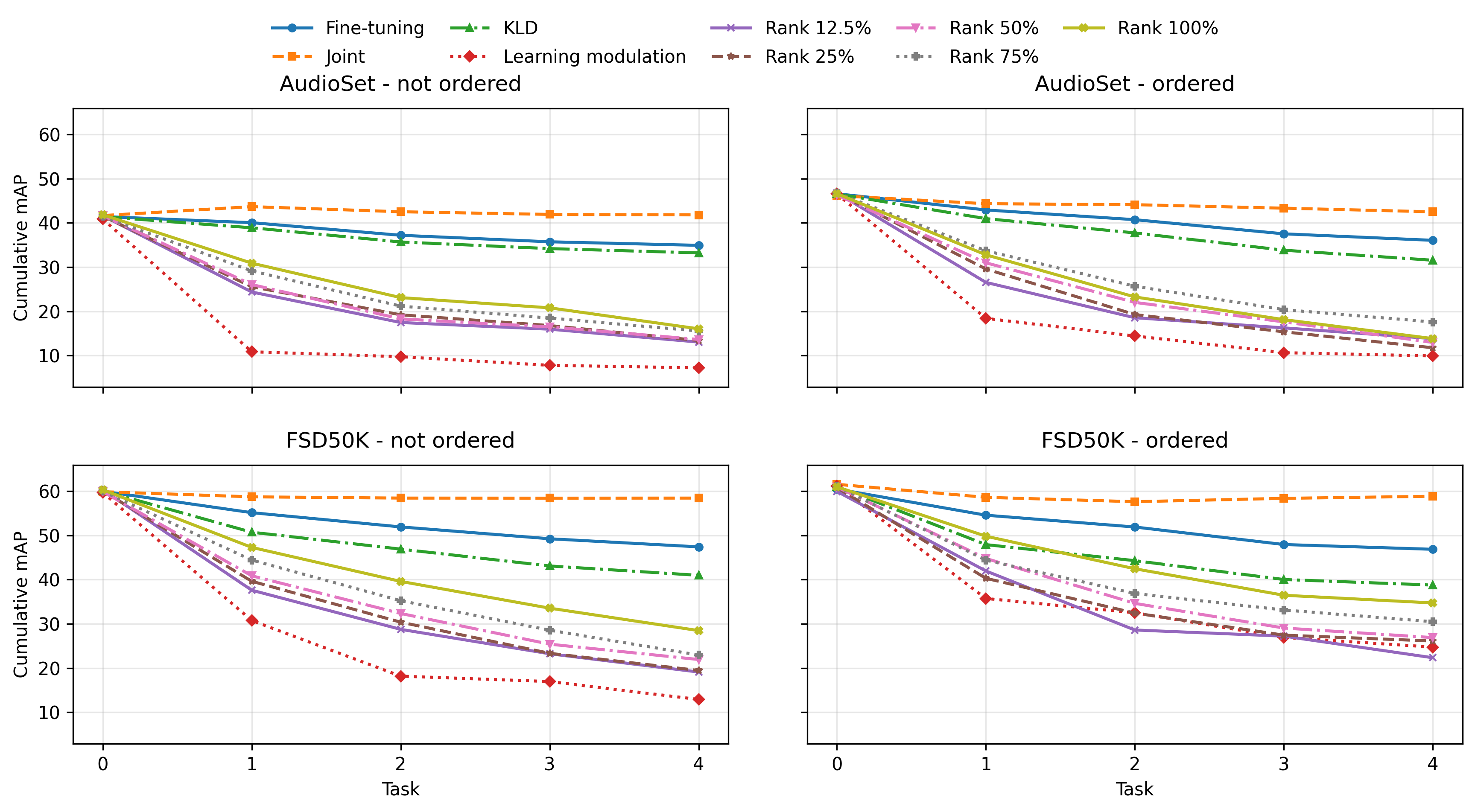}}
\caption{AudioSet and FSD50K Evaluation Metrics for the different learning setups.}
\label{fig:audioset_eval}
\end{figure*}

The performance of the system in the different learning conditions is shown in Figure \ref{fig:audioset_eval}. Performance is expressed using the cumulative mAP after each task, which represents the mAP obtained by the system on all the classes learned so far, i.e. mAP over 30 classes at task 0 (initial task=30), mAP over 35 classes at task 1 (30+5 for tasks 1 and 2), and so on. Based on the evolution of the performance observed in the figures, we make the following observations.

Among the continual learning strategies, fine-tuning and KLD obtain the most competitive results. Fine-tuning is particularly strong in terms of stability, showing only a moderate degradation across tasks and consistently outperforming the rank-based modulation strategies. KLD follows a similar trend, although with a larger performance drop in several configurations. This suggests that the feature extractor retains a substantial amount of useful acoustic knowledge during the incremental stages, and that catastrophic degradation is not primarily caused by a loss of the learned representation.

\begin{table}%[ht]
\centering
\caption{Continual learning metric averages on AudioSet across 5 tasks, for both the "not ordered" dataset configuration (n.o) and the "ordered" dataset configuration (o.).}
\label{tab:cl_audioset}
\begin{tabular}{|l|rr|rr|rr|}
\hline
\textbf{CL solution} & \multicolumn{2}{c|}{\textbf{BWT} $\uparrow$} & \multicolumn{2}{c|}{\textbf{FM} $\downarrow$} & \multicolumn{2}{c|}{\textbf{IM} $\downarrow$} \\
& n.o.& o. & n.o.& o. & n.o.& o. \\
\hline
finetune  & \textbf{-0.5}& \textbf{-0.2}  & 0.2& \textbf{0.1}  & \textbf{5.5}& \textbf{5.2}  \\
kld       & -3.5& -4.7  & 1.1& 1.4  & 6.9& 8.5  \\
learning-modulation  & -16.6& -16.7    & 13.8& 13.1    & 33.5& 31.1   \\
rank 12.5\% & -17.2& -17.6 & 10.2& 10.8 & 24.7& 25.7 \\
rank 25\% & -16.6& -17.7 & 9.5& 10.0  & 23.7& 25.5 \\
rank 50\% & -16.6& -16.2 & 9.4& 8.7  & 23.8& 23.6 \\
rank 75\% & -14.8& -13.9 & 7.7& 7.1  & 21.4& 20.1 \\
rank 100\% & -13.8& -15.6 & 6.9& 8.1  & 19.7& 22.5 \\
\hline
\end{tabular}
\end{table}

The continual learning metrics presented in Tables~\ref{tab:cl_audioset} and~\ref{tab:cl_fsd50k} support this interpretation. On AudioSet, fine-tuning has BWT values close to zero, while KLD introduces slightly stronger forgetting. In contrast, the rank-based strategies and the learning-modulation produce substantially more negative BWT values and higher FM and IM values. This indicates that the rank-based and learning-modulation methods introduce more instability across tasks rather than improving retention. A similar pattern appears on FSD50K, where fine-tuning again has the lowest forgetting among the continual learning methods, while KLD and the other approaches show larger degradation.

\begin{table}%[ht]
\centering
\caption{Continual learning metric averages on FSD50K across 5 tasks, for both the "not ordered" dataset configuration (n.o) and the "ordered" dataset configuration (o.). }
\label{tab:cl_fsd50k}
\begin{tabular}{|l|rr|rr|rr|}
\hline
\textbf{CL solution} & \multicolumn{2}{c|}{\textbf{BWT} $\uparrow$} & \multicolumn{2}{c|}{\textbf{FM} $\downarrow$} & \multicolumn{2}{c|}{\textbf{IM} $\downarrow$} \\
& n.o.& o. & n.o.& o. & n.o.& o. \\
\hline
finetune             & \textbf{-0.5}& \textbf{-0.5}   & \textbf{0.2}& \textbf{0.2}   & \textbf{7.9}& \textbf{8.7}   \\
kld                  & -6.8& -9.3   & 2.3& 4.0   & 13.4& 16.3 \\
learning-modulation  & -19.9& -11.5    & 13.3& 9.0    & 39.2& 29.1   \\
rank 12.5\%          & -21.8& -18.9 & 11.8& 9.7  & 31.7&29.0 \\
rank 25\%            & -20.9& -17.7 & 10.8& 9.6  & 30.7& 27.5 \\
rank 50\%            & -19.0& -15.8 & 9.7& 7.7   & 28.7& 25.2 \\
rank 75\%            & -17.0& -13.4 & 7.8&6.9   & 26.1& 22.8 \\
rank 100\%           & -13.3& -9.8  & 5.6& 3.8   & 21.6& 18.2 \\
\hline
\end{tabular}
\end{table}

The rank-based approaches show a clear sensitivity to the amount of constrained plasticity. More restrictive configurations generally perform better, while less restrictive ones tend to be less stable. For example, the variants that allow a larger portion of the network to remain adaptable usually obtain worse BWT and IM values than the more aggressive rank-based constraints. This suggests that strongly limiting plasticity in the feature extractor is beneficial in this setting. Instead of allowing useful representations to be learned, looser constraints may prevent the model from adapting effectively to the new classes.

These results suggest that the main bottleneck is not necessarily the evolution of the feature extractor, but rather the organization of the output space. The strong performance of the joint reference indicates that, when all classes can be represented in a shared output structure, the model can maintain high performance across tasks. At the same time, the limited benefit of feature-level modulation suggests that the learned representation does not need to change drastically during the incremental stages. The most important factor appears to be providing an appropriate and specific output for the newly introduced classes, reducing interference at the classifier level.

Overall, the experiments indicate that complex feature-level continual learning mechanisms are not clearly advantageous in this scenario. The feature extractor appears to learn general acoustic representations that remain useful across the incremental sequence. The degradation observed in the continual learning setting is therefore more likely related to classifier-level interference and the lack of a suitable output structure for the new classes. From this perspective, the most efficient direction is to focus on output-level organization rather than heavily modifying or constraining the feature extractor during incremental learning.

The above observations are generally valid for both the ordered and not ordered experimental setups. For FSD50K the differences in the CL metrics between the ordered and not ordered learning, with all three indicating that ordered learning is more beneficial. One potential cause for this difference is that FSD50K data, though officially multi-label, contains a vast majority of single-labeled examples, hence the initial task of learning the largest 30 classes provides the best available representation over all future tasks.

\section{Conclusions}
This work investigated a collection of solutions to mitigate catastrophic forgetting for incremental sound classification.
Among the architectural approaches, the dynamic head solution seems to be the solution that helped the most with the catastrophic forgetting, while any kind of regularization approach impacts negatively both the forgetting and the convergence of the networks.
The findings suggest that catastrophic forgetting mainly happens in the classifier head and that the features learned during past tasks can be used by future tasks without the need for any kind of modulation. Thus, we can assume that the feature extractor is not where the catastrophic forgetting is happening for tasks that follow the same probability distribution. 
In future work, we will build on the insights gained here to better understand what happens  for differing class distributions, and to develop effective approaches for alleviating catastrophic forgetting.
\bibliographystyle{IEEEtran}
\bibliography{refs}

\end{document}